\documentstyle [12pt] {article}
\begin{document}

\pagestyle{empty}
\rightline{\vbox{
\halign{&#\hfil\cr
&NUHEP-TH-94-26 \cr
&November 1994 \cr
&hep-ph/9411365 \cr}}}
\bigskip
\bigskip
\bigskip
{\Large\bf
	\centerline{Color-Octet Fragmentation}
	\centerline{and the $\psi'$ Surplus at the Tevatron}}
\bigskip
\normalsize

\centerline{Eric Braaten and Sean Fleming}
\centerline{\sl Department of Physics and Astronomy, Northwestern University,
    Evanston, IL 60208}
\bigskip

\begin{abstract}
The production rate of prompt $\psi'$'s at large transverse momentum
at the Tevatron is larger than theoretical expectations by about a
factor of 30.  As a solution to this puzzle, we suggest that the
dominant $\psi'$ production mechanism is the fragmentation of a gluon
into a $c \bar c$ pair in a pointlike color-octet S-wave state,
which subsequently evolves nonperturbatively into a $\psi'$ plus
light hadrons.  The contribution to the fragmentation
function from this process is enhanced by a short-distance factor of
$1/\alpha_s^2$ relative to the conventional color-singlet contribution.
This may compensate for the suppression by $v^4$, where $v$ is the
relative momentum of the charm quark in the $\psi'$.  If this is indeed
the dominant production mechanism at large $p_T$, then the prompt $\psi'$'s
that are observed at the Tevatron should almost always be associated
with a jet of light hadrons.
\end{abstract}

\vfill\eject\pagestyle{plain}\setcounter{page}{1}

The production of heavy quarkonium in high energy processes provides
severe tests of our understanding of QCD in a domain that should
be accessible to theoretical analysis.
At high energies, QCD mechanisms for the direct production of
charmonium are embedded in a large background from
the production of $b$ quarks, followed by the
decay of the resulting $B$ hadrons into charmonium states.
Vertex detectors can be used to separate the charmonium states
coming from $b$ quarks from those that
are produced by QCD interactions,
thus allowing the direct study of the prompt QCD production mechanisms.
In the 1991-92 run of the Tevatron, the CDF detector group used their
new vertex detector to study prompt
charmonium production~\cite{vaia,troy}.
The results have proved to be an embarrassment to theory, revealing
orders-of-magnitude discrepancies with theoretical predictions.

Until recently, the conventional wisdom was that the dominant
mechanisms for prompt charmonium production at large
transverse momentum $p_T$ come from the
processes such as gluon-gluon fusion which are leading order in
the QCD coupling constant $\alpha_s$~\cite{br}.
The resulting predictions for prompt $J/\psi$ production at large $p_T$
are smaller than the CDF data by more than an order of magnitude.
However, as pointed out by Braaten and Yuan in 1990 \cite{by}, the dominant
production mechanism at sufficiently large $p_T$ is the production
of a parton with large transverse momentum, followed by the fragmentation
of the parton into charmonium states.  Including this fragmentation
mechanism brings the theoretical predictions for prompt $J/\psi$ production
to within a factor of 3 of the data \cite{bdfm,cg}, close enough that the
remaining discrepancy can be attributed to theoretical uncertainties.
Unfortunately, even after including the fragmentation contribution, the
prediction for the $\psi'$ production rate remains a factor of 30 below
the data.  This remarkably large discrepancy between theory and
experiment suggests that an entirely new production mechanism must be
dominating the production of $\psi'$ at large $p_T$.

Previous proposals for solving the $\psi'$ surplus problem have focused
on the production of higher charmonium states which decay into $\psi'$.
Among the possible states are higher P-wave states, higher D-wave states,
and states whose dominant Fock component
is $|c \bar c g \rangle$~\cite{close}.
For each of these possibilities, there is reason to question whether
the production rate can be large enough and whether the branching fraction
into $\psi'$ can be large enough to account for the observed rate for
prompt $\psi'$ production.
In this Letter, we propose a completely different solution to the
$\psi'$ surplus problem.  We suggest that the dominant production mechanism
for $\psi'$ at large $p_T$ is gluon fragmentation into a $c \bar c$ pair
in a pointlike color-octet S-wave state,
followed by the nonperturbative evolution of
this state into a $\psi'$ plus light hadrons.  We calculate the
corresponding contribution to the fragmentation function for
$g \rightarrow \psi'$ in terms of a well-defined nonperturbative
matrix element, and we determine the value of this matrix element
that would be required to explain the production rate of $\psi'$ at the
Tevatron.  We then discuss possible tests of this solution to the
$\psi'$ surplus problem.

Our proposal is based on a
general factorization analysis of the annihilation and production
of heavy quarkonium by Bodwin, Braaten, and Lepage \cite{bbl}.
The factorization formalism of
Ref.~\cite{bbl} allows the fragmentation functions for charmonium production
to be factored into short-distance
coefficients, which describe the production rate of a $c \bar c$ pair
within a region of size $1/m_c$, and long-distance factors that
contain all the nonperturbative dynamics of the formation
of a bound state containing the $c \bar c$ pair.  The fragmentation function
for a gluon to split into a charmonium state $X$ with longitudinal
momentum fraction $z$ can be written
\begin{equation}
D_{g \rightarrow X}(z,\mu)
\;=\; \sum_n d_n(z,\mu) \; \langle 0| {\cal O}^X_n |0 \rangle ,
\label{D}
\end{equation}
where ${\cal O}^X_n$ are local 4-fermon operators that were defined
in Ref.~\cite{bbl} in terms of the fields of nonrelativistic QCD~\cite{NRQCD}.
The short-distance coefficients $d_n(z,\mu)$ are independent of
the quarkonium state $X$.
For a fragmentation scale $\mu$ of order $m_c$, they
can be computed using perturbation theory in $\alpha_s(m_c)$.
The dependence on the quarkonium state $X$ appears in the
long-distance factors $\langle {\cal O}^X_n \rangle$.
The relative importance of the various matrix elements
for a given state $X$ can be estimated
by how they scale with $v$, the typical relative velocity of the charm quark
in charmonium \cite{bbl}.

In previous calculations of fragmentation functions for
charmonium production,  only the matrix elements that dominate in the
nonrelativistic limit $v \rightarrow 0$ have been considered.
In the case of $\psi'$, the only such matrix element
in the notation of Ref.~\cite{bbl} is
$\langle {\cal O}^{\psi'}_1(^3S_1) \rangle$.  In this vacuum matrix element,
the local operator ${\cal O}^{\psi'}_1(^3S_1)$
creates a $c \bar c$ pair in a color-singlet $^3S_1$ state,
projects onto the subspace of states which in the asymptotic future contain
a $\psi'$, and then annihilates the $c \bar c$ at the creation point.
Up to corrections of relative order $v^4$, this matrix element can be
related to the wavefunction $\overline{R_{\psi'}}/\sqrt{4 \pi}$
of the $\psi'$ at the origin:
\begin{equation}
\langle 0| {\cal O}^{\psi'}_1(^3S_1) |0 \rangle
\;\approx\; {3 \over 2 \pi} |\overline{R_{\psi'}}|^2 .
\end{equation}
In the case of S-waves, the factorization approach reduces in the
nonrelativistic limit to the color-singlet model~\cite{schuler},
where the $\psi'$ is treated as a $c \bar c$ pair in a color-singlet
$^3S_1$ state with vanishing relative momentum.

In the case of P-waves like the $\chi_{cJ}$ state, there are 2 matrix
elements that survive in the nonrelativistic limit.  In addition to
$\langle {\cal O}^{\chi_{cJ}}_1(^3P_J) \rangle$, which is related to the
derivative of the wavefunction at the origin, there is also the
color-octet matrix element $\langle {\cal O}^{\chi_{cJ}}_8(^3S_1) \rangle$.
The local operator ${\cal O}^{\chi_{cJ}}_8(^3S_1)$
creates a $c \bar c$ pair in a color-octet $^3S_1$ state,
projects onto the subspace of states which in the asymptotic future contain
a $\chi_{cJ}$, and then annihilates the $c \bar c$ at the creation point.
In the color-singlet model, only the matrix element
$\langle {\cal O}^{\chi_{cJ}}_1(^3P_J) \rangle$ is taken into account.
This model is perturbatively inconsistent because of infrared divergences
in the radiative corrections to P-wave production rates.
In the factorization formalism of Ref.~\cite{bbl}, this problem is solved
by factoring the infrared divergences into the color-octet matrix element
$\langle {\cal O}^{\chi_{cJ}}_8(^3S_1) \rangle$~\cite{bbly}.

The restriction to the leading matrix elements in the $v \rightarrow 0$ limit
is a great simplification, but it is not necessarily correct.
The typical value of $v^2$ in the case of charmonium is only about
$1/3$, so factors of $v^2$ do not provide a very large suppression.
In the case of (\ref{D}), dropping contributions that are suppressed
by powers of $v^2$ might be reasonable if the short-distance factors
$d_n(z,\mu)$ were all comparable in size.  However these factors are of
different order in $\alpha_s$ for different matrix elements.
Short-distance factors of $\alpha_s(m_c)/\pi$ probably provide much more
effective suppression than factors of $v^2$.
Thus, in estimating the importance of various
terms in the fragmentation function, one should take into account not
only the scaling of the matrix element with $v^2$, but also the order
in $\alpha_s$ of the short-distance coefficient.

In the case of gluon fragmentation into $\psi'$, the
short-distance factor associated with the matrix element
$\langle {\cal O}^{\psi'}_1(^3S_1) \rangle$ is of order $\alpha_s^3$.
It describes the process $g^* \rightarrow c \bar c g g$,
with the $c \bar c$ pair in a color-singlet $^3S_1$ state.
By the velocity-scaling rules of Ref.~\cite{bbl},
the matrix element $\langle {\cal O}^{\psi'}_1(^3S_1) \rangle$
is of order $m_c^3 v^3$, so
the contribution to the fragmentation function
is of order $\alpha_s^3 v^3$.  While all other matrix elements are
suppressed by powers of $v^2$, there are some with fewer short-distance
suppression factors of $\alpha_s$.  In particular, there are some matrix
elements for which the leading-order
short-distance process is $g^* \rightarrow c \bar c$
and the rate is of order $\alpha_s$.
Of these matrix elements, the leading one in the nonrelativistic limit is
$\langle {\cal O}^{\psi'}_8(^3S_1) \rangle$, which scales like $m_c^3v^7$.
The suppression factor of $v^4$ relative to
$\langle {\cal O}^{\psi'}_1(^3S_1) \rangle$ reflects the fact that
the $c \bar c$ pair in the $\psi'$ can make a double E1 transition
with amplitude of order $v^2$ from the dominant
$|c \bar c \rangle$ Fock state to a $|c \bar c g g \rangle$ state
in which the $c \bar c$ pair is in a color-octet $^3S_1$ state.

The contribution to the fragmentation function for $g \rightarrow \psi'$
from the short-distance production of a $c \bar c$ pair in a color-octet
$^3S_1$ state was calculated to leading order in $\alpha_s$ in Ref.~\cite{byb}.
The calculation was carried out for the specific case of
$g \rightarrow \chi_{cJ}$, but since the short-distance coefficients
in (\ref{D}) are independent of the quarkonium state, the result applies
equally well to $\psi'$.  The fragmentation function is
\begin{equation}
D_{g \rightarrow \psi'}(z,\mu)
\;=\; {\pi \alpha_s(2 m_c) \over 24 m_c^3} \delta(1-z)
	\; \langle 0| {\cal O}^{\psi'}_8(^3S_1) |0 \rangle .
\label{Dpsi}
\end{equation}
In Ref.~\cite{byb}, the corresponding fragmentation function for
$g \rightarrow \chi_{cJ}$ was expressed in terms of a quantity
$H_8'$ defined by
\begin{equation}
H_8' \;\equiv\; {1 \over (2J+1) m_c^2}
\langle 0| {\cal O}^{\chi_{cJ}}_8(^3S_1) |0 \rangle .
\end{equation}
The form (\ref{Dpsi}) is preferred, since it corresponds to the
factorization of the fragmentation  function into a short-distance factor
proportional to $\alpha_s/m_c^3$ and a long-distance matrix element.

In the fragmentation function for $g \rightarrow \chi_{cJ}$,
the matrix element $\langle {\cal O}^{\chi_{cJ}}_8(^3S_1) \rangle$
can be determined phenomenologically from data on charmonium production
in $B$-meson decays.  From eq. (24) of Ref.~\cite{bbly},
we have
\begin{equation}
{\Gamma(B \rightarrow \chi_{c2} + X)
	\over \Gamma(B \rightarrow e {\bar \nu}_e + X)}
\;\approx\; 14.8 \; { \langle 0| {\cal O}^{\chi_{c2}}_8(^3S_1) |0 \rangle
	\over m_c^2 m_b}.
\label{Bchi}
\end{equation}
The branching fraction for $B \rightarrow \chi_{c2} + X$ has recently been
measured by the CLEO collaboration to be $(0.25 \pm 0.11) \%$ \cite{cleo}.
Assuming
$m_c \approx 1.5$ GeV and $m_b \approx 4.5$ GeV, the expression
(\ref{Bchi}) gives
\begin{equation}
\langle 0| {\cal O}^{\chi_{c2}}_8(^3S_1) |0 \rangle
\;=\; (0.016 \pm 0.007) \; {\rm GeV}^3 ,
\label{O8chi}
\end{equation}
which corresponds to $H_8' \approx (1.4 \pm 0.6)$ MeV. This value is
significantly smaller than the estimate $H_8' \approx 3$ MeV
obtained in Ref.~\cite{bbly} and used in
Refs.~\cite{bdfm} and \cite{cg}.  Thus the contribution
to $J/\psi$ production from gluon fragmentation into $\chi_{cJ}$
that was calculated in Refs.~\cite{bdfm} and \cite{cg}
should be decreased by a factor of 2.  The theoretical prediction
for $J/\psi$ production then falls below the data by about a factor of 5.

In the case of $\psi'$, we can determine the matrix element
$\langle {\cal O}^{\psi'}_8(^3S_1) \rangle$ by fitting the
CDF data on the production rate of prompt $\psi'$ at large $p_T$.
The production rate is given by convoluting hard scattering cross sections
for the production of gluons with large transverse momentum with
parton distributions for the colliding proton and antiproton and with
the fragmentation function $D_{g \rightarrow \psi'}(z,\mu)$.
The fragmentation function given by (\ref{Dpsi}) is evolved up to the scale
$\mu = p_T/z$ of the transverse momentum of the fragmenting gluon by
using the Altarelli-Parisi equations.
The production rate measured by CDF~\cite{troy} in the 1991-92 run is shown in
Figure 1, along with three theoretical curves.  The dotted line is
the production rate from leading-order production mechanisms, such as
$g g \rightarrow \psi' g$, while the dashed line is the color-singlet
fragmentation contribution calculated in Ref.~\cite{bdfm},
which is dominated by charm quark fragmentation into $\psi'$.
The solid curve is the color-octet
fragmentation contribution, with the matrix element
$\langle {\cal O}^{\psi'}_8(^3S_1) \rangle$ adjusted to make the
normalization agree with the data.  The resulting value of the
matrix element is
\begin{equation}
\langle 0| {\cal O}^{\psi'}_8(^3S_1) |0 \rangle
\;=\; 0.0042 \; {\rm GeV}^3 .
\label{O8}
\end{equation}
Note that the shape of the color-octet fragmentation contribution
fits the data perfectly.

There is a very simple consistency check that the color-octet
fragmentation mechanism must satisfy in order to provide a viable
explanation for the $\psi'$ surplus.  The value of the matrix
element $\langle {\cal O}^{\psi'}_8(^3S_1) \rangle$ must be small
enough relative to the matrix element
$\langle {\cal O}^{\psi'}_1(^3S_1) \rangle$ to be consistent
with suppression by a factor of $v^4$. The color-singlet matrix element
can be determined from the leptonic decay
rate of the $\psi'$ to be $0.11 \; {\rm GeV}^3$.  The color-octet
matrix element (\ref{O8}) is smaller by a factor of 25,
consistent with suppression by $v^4$.

There are several experimental consistency checks of
our proposed explanation of the $\psi'$ surplus.
If color-octet fragmentation is indeed the dominant
production mechanism, then most of the $\psi'$'s
produced at the Tevatron must be accompanied by light
hadrons.  The leading order mechanisms, such as $g g \rightarrow \psi' g$,
produce isolated $\psi'$'s with no accompanying hadrons.
Color-singlet fragmentation produces $\psi'$'s that are associated with a
charm quark jet, but with transverse momentum of order $m_c$ relative
to the jet.
Color-octet fragmentation also produces $\psi'$'s
that are associated with a jet of light
hadrons, but the transverse momentum relative to the jet is of order
$m_c v$ or smaller.  Thus measurements of the hadronic energy distribution
as a function of the distance from the $\psi'$ momentum in azimuthal angle
and in rapidity could provide evidence for this new production mechanism.
It has recently been pointed out by Cho and Wise~\cite{cho}
that measurements of the spin alignment of the $\psi'$'s, which
is reflected in the angular distribution of their leptonic decays,
might provide a clear signature for the color-octet fragmentation
mechanism.  This mechanism produces $\psi'$'s that are 100\%
transversely polarized at leading order in $\alpha_s$, while other mechanisms
tend to produce unpolarized $\psi'$'s.

If color-octet fragmentation is so important in the case of $\psi'$
production, one might ask why it is not equally important for
$J/\psi$ production at large $p_T$.  In fact, it might very well be.
The main difference is that the $J/\psi$ signal is fed by
gluon fragmentation into $\chi_{c1}$ and $\chi_{c2}$ followed by
their radiative decays into $J/\psi$,
and this overwhelms the rate for color-singlet
fragmentation directly into  $J/\psi$.
However there is room for a substantial contribution from color-octet
fragmentation into $J/\psi$, especially with the new determination
(\ref{O8chi}) of the matrix element
$\langle {\cal O}^{\chi_{c2}}_8(^3S_1) \rangle$, which leaves the
predictions of  Refs.~\cite{bdfm} and \cite{cg} a factor of 5 below
the CDF data.  If the matrix element
$\langle {\cal O}^{J/\psi}_8(^3S_1) \rangle$ for $J/\psi$ has a value
comparable to that for $\psi'$ given in (\ref{O8}),
then the color-octet fragmentation mechanism for $J/\psi$ production
will be comparable in importance to gluon fragmentation into
$\chi_{c1}$ and $\chi_{c2}$.

If a color-octet contribution dominates the fragmentation function
for $g \rightarrow \psi'$, one might worry that annihilation of a
$c \bar c$ pair in a color-octet state should also
dominate the decay rate of the $\psi'$ into light hadrons.
However the relative importance of the color-singlet
and color-octet contributions is different in the case of
annihilation.  The color-singlet contribution to the decay rate is of
order $\alpha_s^3 m_c v^3$, while annihilation from a color-octet
S-wave state contributes at order $\alpha_s^2 m_c v^7$.  Thus the
short-distance enhancement of the color-octet contribution is only
one power of $1/\alpha_s$, rather than 2 powers as in the case of
fragmentation.  Furthermore, there is no apparent rigorous relation
between the color-octet annihilation matrix elements and the color-octet
production matrix elements beyond the fact that they have the same
scaling in $v$ \cite{bbl}.

Our proposal for the solution of the $\psi'$ surplus problem
has predictive power, because the matrix element
$\langle {\cal O}^{\psi'}_8(^3S_1) \rangle$ determined from $\psi'$
production at large $p_T$ can be used to predict
$\psi'$ production in other  high energy processes.  Unfortunately,
the matrix elements $\langle {\cal O}^{\psi'}_8(^1S_0) \rangle$
and $\langle {\cal O}^{\psi'}_8(^3P_J) \rangle$ are of the same order
in $v^2$ as $\langle {\cal O}^{\psi'}_8(^3S_1) \rangle$,
and they must also be taken into account.
Different production processes probe different linear combinations of
these 3 color-octet matrix elements at leading order in $\alpha_s$.
Gluon fragmentation is rather unique in that
$\langle {\cal O}^{\psi'}_8(^3S_1) \rangle$ is the only color-octet
matrix element that contributes at leading order in $\alpha_s$ and in $v^2$.
In photoproduction of $\psi'$, on the other hand,
only the matrix elements $\langle {\cal O}^{\psi'}_8(^1S_0) \rangle$
and $\langle {\cal O}^{\psi'}_8(^3P_J) \rangle$ contribute
at leading order in $\alpha_s$ and in $v^2$.  While suppressed
relative to the color-singlet contribution by a factor of $v^4$,
they are enhanced by a short-distance factor of $1/\alpha_s$.
Moreover, at leading order in $\alpha_s$, the color-octet contribution
is concentrated near $z = 1$, where $z$ is the ratio of the energy of the
$\psi'$ to that of the photon. The cross section is known experimentally
to peak strongly in the $z \rightarrow 1$ region, and charmonium states
produced in this region are labelled ``elastic''.  Thus a linear
combination of the matrix elements $\langle {\cal O}^{\psi'}_8(^1S_0) \rangle$
and $\langle {\cal O}^{\psi'}_8(^3P_J) \rangle$ can be determined from the
elastic peak in $\psi'$ photoproduction \cite{bf}.

In most of the work on the production of charmonium over the last decade,
the dominant production mechanisms have been assumed to be ones in which the
$c \bar c$ pair is produced at short distances in a color-singlet state.
However, in the case of P-wave states,
it is essential to include the color-octet
mechanism to obtain consistent perturbative predictions.
In the case of S-waves, the color-octet mechanism is not essential for
perturbative consistency, but it may be important nonetheless in cases where
the color-singlet mechanism is suppressed by a short-distance coefficient.
A global analysis of the data on $\psi'$ production from all high energy
processes could confirm the importance of color-octet mechanisms
in the production of the S-wave states of heavy quarkonium.

This work was supported in part by the U.S. Department of Energy,
Division of High Energy Physics, under Grant DE-FG02-91-ER40684.

\bigskip
\noindent{\Large\bf Figure Captions}
\begin{enumerate}
\item CDF data for prompt $\psi'$ production compared with theoretical
calculations of the total leading order contributions (dotted curves),
the total color-singlet fragmentation contributions (dashed curves),
and the color-octet fragmentation contribution (solid curves).
The normalization of the color-octet fragmentation contribution
has been adjusted to fit the data.
\end{enumerate}
\vfill\eject


\bigskip
\begin{thebibliography}{99}

\bibitem{vaia}
V. Papadimitriou, CDF Coll., presented at the ``Rencontres de la Vallee
        d'Aoste'', La Thuile, March 1994.

\bibitem{troy}
T. Daniels, CDF Coll., Fermilab-Conf-94/136-E.

\bibitem{br}
R. Baier and R. R\"{u}ckl, Z. Phys. {\bf C19} 251 (1983);
F. Halzen, F. Herzog, E.W.N. Glover, and A.D. Martin,
	Phys. Rev. {\bf D30}, 700 (1984);
E.W.N. Glover, A.D. Martin, W.J. Stirling, Z. Phys. {\bf C38}, 473 (1988);
B. van Eijk and R. Kinnunen, Z. Phys. {\bf C41}, 489 (1988).

\bibitem{by}
E. Braaten and T.C. Yuan, Phys. Rev. Lett. {\bf 71}, 1673 (1993).

\bibitem{bdfm}
E. Braaten, M.A. Doncheski, S. Fleming, and M. Mangano,
	Phys. Lett. {\bf B333}, 548 (1994).

\bibitem{cg}
M. Cacciari and M. Greco, Phys. Rev. Lett. {\bf 73}, 1586 (1994);
D.P. Roy and K. Sridhar, CERN-TH.7329/94 (hep-ph 9406386).

\bibitem{close}
P. Cho, S. Trivedi and M. Wise, CALT-68-1943 (hep-ph  9408352);
F.E. Close, RAL-94-093 (hep-ph  9409203);
D.P. Roy and K. Sridhar, CERN-TH.7434/94 (hep-ph  9409232);
P. Cho and M. Wise, CALT-68-1954 (hep-ph 9410214).

\bibitem{bbl}
G.T. Bodwin, E. Braaten, and G.P. Lepage, Argonne preprint
    ANL-HEP-PR-94-24 (hep-ph  9407339).

\bibitem{NRQCD}
W.E. Caswell and G.P. Lepage, Phys. Lett. {\bf 167B}, 437 (1986).

\bibitem{schuler} See, for example, G.A. Schuler, CERN preprint
    CERN-TH.7170/94 (hep-ph 9403387), and references therein.

\bibitem{bbly}
G.T. Bodwin, E. Braaten, T.C. Yuan, and G.P. Lepage,
	Phys. Rev. {\bf D46}, R3703 (1992).

\bibitem{byb}
E. Braaten and T.C. Yuan, Phys. Rev. {\bf D50}, 3176 (1994).

\bibitem{cleo}
CLEO Collaboration report CLEO CONF 94-11, submitted to the
    Int. Conf. on High Energy Physics, Glasgow,
    July 1994 (Ref. GLS0248).

\bibitem{cho}
P. Cho and M. Wise, CALT-68-1962 (hep-ph 9411303).

\bibitem{bf}
E. Braaten and S. Fleming (in progress).

\end{thebibliography}
\end{document}